\documentclass[preprint,showpacs,preprintnumbers,superscriptaddress,amsmath,amssymb]{revtex4}

\usepackage{graphicx}
\usepackage{dcolumn}
\usepackage{bm}
\usepackage[latin1]{inputenc}
\usepackage{float}
\usepackage{amsmath}
\usepackage{amssymb}

\begin{document}


\title{Frequency-Domain Measurement of the Spin Imbalance Lifetime in Superconductors}

\author{C. H. L. Quay}
 \email{charis.quay@u-psud.fr}
 \affiliation{%
Laboratoire de Physique des Solides (CNRS UMR 8502), Bâtiment 510, Université Paris-Sud, 91405 Orsay, France
}%

\author{C. Dutreix}
 \affiliation{%
Laboratoire de Physique des Solides (CNRS UMR 8502), Bâtiment 510, Université Paris-Sud, 91405 Orsay, France
}%
\author{D. Chevallier}
 \affiliation{%
Department of Physics, University of Basel, Klingelbergstrasse 82, CH-4056 Basel, Switzerland
}%
\author{C. Bena}
 \affiliation{%
Laboratoire de Physique des Solides (CNRS UMR 8502), Bâtiment 510, Université Paris-Sud, 91405 Orsay, France
}%
 \affiliation{%
Institut de Physique Théorique, CEA Saclay 91190 Gif-sur-Yvette, France.
}%

\author{M. Aprili}

 \affiliation{%
Laboratoire de Physique des Solides (CNRS UMR 8502), Bâtiment 510, Université Paris-Sud, 91405 Orsay, France
}%

\date{\today}

\begin{abstract}
We have measured the lifetime of spin imbalances in the quasiparticle population of a superconductor ($\tau_s$) in the frequency domain. A time-dependent spin imbalance is created by injecting spin-polarised electrons at finite excitation frequencies into a thin-film mesoscopic superconductor (Al) in an in-plane magnetic field (in the Pauli limit). The time-averaged value of the spin imbalance signal as a function of excitation frequency, $f_{RF}$ shows a cut-off at $f_{RF} \approx 1/(2\pi\tau_s)$. The spin imbalance lifetime is relatively constant in the accessible ranges of temperatures, with perhaps a slight increase with increasing magnetic field. Taking into account sample thickness effects, $\tau_s$ is consistent with previous measurements and of the order of the electron-electron scattering time $\tau_{ee}$. Our data are qualitatively well-described by a theoretical model taking into account all quasiparticle tunnelling processes from a normal metal into a superconductor.
\end{abstract}


\maketitle

Spin-polarised electrons injected into superconductors eventually disappear into the condensate, which is made up of Cooper pairs of electrons of opposite spin. To disappear, the injected electrons --- which become quasiparticles in the superconductor --- must lose energy, flip their spin and recombine with quasiparticles of the opposite spin to form Cooper pairs. These processes may be sequential or occur in parallel. For example, (1) quasiparticles may undergo elastic or inelastic spin flip processes, (2) quasiparticles may lose energy without flipping their spin, and (3) low-energy quasiparticles recombining in pairs necessarily lose a quantity of energy equivalent to the superconducting gap, usually in the form of a phonon. The characteristic timescale for these processes --- as well as the order in which they occur and any interdependence between them --- can shed light on microscopic properties of the superconductor, including relaxation pathways~\cite{schrieffer,rothwarf,owen,chang,devereaux,reizer,quinlan,yafet,wakamura} as well as the gap structure and the pairing mechanism in unconventional superconductors~\cite{averitt,demsar,gedik,madan}.

Time- and frequency-domain experiments, whether using transport, optical pump-probe or other techniques, provide the most direct measure of the timescales involved~\cite{johnson,carr,peters,hu}. Most of the work in this area has focused on the recombination of quasiparticles, usually with techniques sensitive to the number of quasiparticles and their diminution over time. A quasiparticle population which is larger than that at equilibrium does not however exhaust the possible non-equilibrium phenomena: the quasiparticle population can also manifest spin and/or charge imbalances~\cite{clarke,tinkham,quay,hubler,wolf,zhao,takahashi}. These do not necessarily relax in the same way, nor on the same timescale. Here we report the first frequency-domain measurement of the lifetime of a spin imbalance in the quasiparticle population in a mesoscopic superconductor.

The idea of our experiment is as follows: We inject spin-polarised quasiparticles into a superconductor, in Zeeman field, at a finite frequency $f_{RF} = \omega/2\pi$ while measuring the time-average of the non-local signal due to the resulting spin imbalance $S(\omega,t)$. We expect a cut-off at $\omega = \alpha/\tau_s$, with $\tau_s$ the spin lifetime of quasiparticles in the superconductor and $\alpha$ a constant; as explicated below, this is visible because of the highly non-linear current-voltage characteristic of our detector.

\begin{figure}[!h]
\centering
\includegraphics[width=8.6cm]{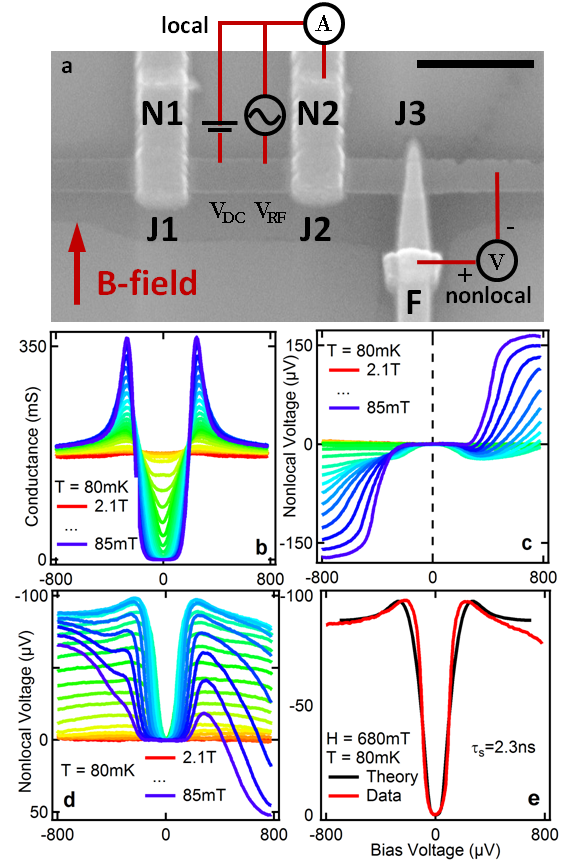}
\caption{\label{fig:sample} (a) Scanning electron micrograph of a typical device (scale bar = 1$\mu$m) and schematic drawing of the measurement setup. S = superconductor ($\sim$8.5nm thick Al film with a native oxide), N = normal metal (100nm Al), F = ferromagnet (40nm Co, with a 4.5nm Al capping layer). The native oxide on S constitutes a tunnel barrier between it and any other given electrode. Quasiparticles are injected into S across a tunnel barrier by applying a voltage $V_{DC}$ across J1 or J2. These are spin-polarised because of the Zeeman field in S. The non-local voltage $V_{NL}$ and differential non-local signal $dV_{NL}/dV_{DC}$ are measured between F and S (at J3) as a function of magnetic field and temperature, as well as as a function of the amplitude $V_{RF}$ and frequency $f_{RF} = \omega/2\pi$ of high-frequency (1-50MHz) voltages applied to the injection electrode. The local conductance $dI/dV_{DC}$ is measured simultaneously at the injection electrode. (b) The conductance $dI/dV_{DC}$ across J2, which is proportional to the quasiparticle density of states in the superconductor, as a function of $V_{DC}$ at different magnetic fields. (c) The non-local voltage $dV_{NL}$ measured at J1 as a function of $V_{DC}$ at the same fields. (d) The non-local voltage $dV_{NL}$ measured at J3 as a function of $V_{DC}$ at the same fields. (e) Theoretical fit to one of the traces in (b), which yields an estimate of $\tau_s$ at $H$=680mT of 2.3ns. We also obtained 1.2ns, 1.6ns and 2.6ns for 425mT, 510mT and 936mT respectively. The error is 10-20\% based on the fits and could be larger if uncertainties in the spin-resolved DOS are considered.}
\end{figure}

Our samples, fabricated with standard electron-beam lithography and evaporation techniques, are thin-film superconducting (S) bars, with a native insulating (I) oxide layer, across which lie normal metal (N) and ferromagnetic (F) electrodes used either as `injectors' or as `detectors'. (Figure~\ref{fig:sample}) In our devices, S is aluminium (8.5nm), I Al$_2$O$_3$, F cobalt (40nm with an Al capping layer) and N thick aluminium (100nm) with a critical magnetic field of $\sim$50mT. (All the data shown were taken with this Al electrode in the normal state.) A typical device is shown in Figure 1a. As in previous experiments, the SIF and NIS junctions have `area resistances' respectively of $\sim2$ and $\sim6$ $\cdot10^{-6}\Omega\cdot cm^2$ (corresponding to barrier transparencies of $\sim4$ and $\sim1\cdot10^{-5}$) and tunnelling is the main transport mechanism across the insulator. (See Supp. Info. of Ref.~\cite{quay}) Measurements were performed at temperatures down to 50mK, in a dilution refrigerator.

We simultaneously perform local and non-local transport measurements using standard lock-in techniques: We apply a voltage $V_{DC}$ across junction J2, between N and S, and measure the (`local') current $I$ injected into the superconductor through J2 and the (`non-local') voltages across the other junctions (J1 and J3), which act as detectors. We also measure the local conductance $dI/dV_{DC}$ and the non-local differential signal, $dV_{NL}/dV_{DC}$. (The lock-in frequency is typically $\sim$10Hz and the excitation voltage 10--20$\mu$V.) The distance between injection and detection junctions is $\lesssim$1$µ$m, well within the spin imbalance relaxation length in the superconducting state in Al~\cite{hubler}. In the presence of an in-plane magnetic field, $H$ (applied parallel to the non-superconducting electrodes), electrons injected into the superconductor create a spin imbalance in its quasiparticle population due to the Zeeman effect~\cite{quay}. The non-local voltage drop $V_{NL}$ at J3 is proportional to either ($\mu_{QP\uparrow}- \mu_P$) or ($\mu_{QP\downarrow}- \mu_P$), depending on the relative alignments of the F magnetisation and the magnetic field. Here $\mu_{QP\beta}$ is the chemical potential of the spin $\beta$ quasiparticle population and $\mu_P$ the Cooper pair chemical potential. We remind the reader that $\mu_C = (\mu_{QP\uparrow}+\mu_{QP\downarrow})/2$ and $\mu_S = (\mu_{QP\uparrow}-\mu_{QP\downarrow})/2$ quantify charge and spin accumulation respectively. The non-local voltage drop at J1 is proportional to $\mu_C- \mu_P$.

To explore the frequency dependence of the spin imbalance, we add higher-frequency components of amplitude $V_{RF}$ and frequency $f_{RF} = $500kHz--50MHz to $V_{DC}$ via a bias-tee located next to the device and at low-temperature. (See Figure~\ref{fig:sample}a and Supp. Info.)

Before presenting the experimental data, let us sketch out our main theoretical expectations. We assume that the spin accumulation, $S$ in the superconductor satisfies
\begin{equation}
\frac{d S(t)}{dt}=I_s(t)-\frac{S(t)}{\tau_s},
\end{equation}
where $\tau_s$ is the spin relaxation time in the superconductor and $I_s$ the spin current.

This equation admits an exact analytical solution:
\begin{equation} \label{}
S(t) = e^{-t/\tau_s} \int_0^{t} dt'I_s(t') e^{t'/\tau_s}.
\end{equation}

We first consider a spin current of the form $I_s(t) = I_{DC}+ I_{RF}e^{i\omega t}$, where $I_{DC}$ and $I_{RF}$ are constants, we then have
\begin{equation} \label{eq:st}
S(t) = \tau_s I_{DC} + \frac{\tau_s I_{RF}}{1+\omega^2\tau_s^2}e^{i(\omega t + \phi)}+ \text{transient terms}
\end{equation}
with $\phi$ a constant phase. The amplitude of the oscillations in $S(t)$ (and thus $\mu_s(t)$ and $V_{NL}(t)$, the quantity we measure) are frequency-dependent and show a Lorentzian cut-off behaviour; however, the time-averaged spin accumulation $<S(\omega,t)>_t$ is frequency-independent. This would seem to imply that high-frequency detection is required.

Nevertheless, considering a voltage bias and non-linear current-voltage characteristics at injector and/or detector junctions, numerical calculations show that there is also a cut-off in $<V_{NL}(\omega,t)>_t$ at $\omega = \alpha/\tau_s$~\cite{chevallier}. In our devices, both injection and detection junctions are non-linear, due to the energy-dependent Bardeen-Cooper-Schrieffer (BCS) quasiparticle density of states (DOS) in the superconductor. Therefore the high-frequency cut-off of Equation~\ref{eq:st} also appears in the DC non-local voltage, and a DC measurement of $\tau_s$ is possible. According to calculations based on DOS extracted from the measured conductance $dI/dV_{DC}$, $\alpha$ can vary from about 0.2 to 0.6, depending on both $V_{DC}$ and $H$.

Our theoretical model is described in Ref.~\cite{chevallier} and is based on previous work by Zhao and Hershfield~\cite{zhao}, which takes into account all quasiparticle tunnelling process at a normal-superconducting junction, extended to include the Zeeman effect induced by the magnetic field. In contrast to the (similar) model presented in our previous work~\cite{quay}, no assumptions were made about the amplitude of the Zeeman energy or $\mu_S$ (which can be up to half the size of the superconducting gap in these measurements).

\begin{figure}[!h]
\includegraphics[width=8.6cm]{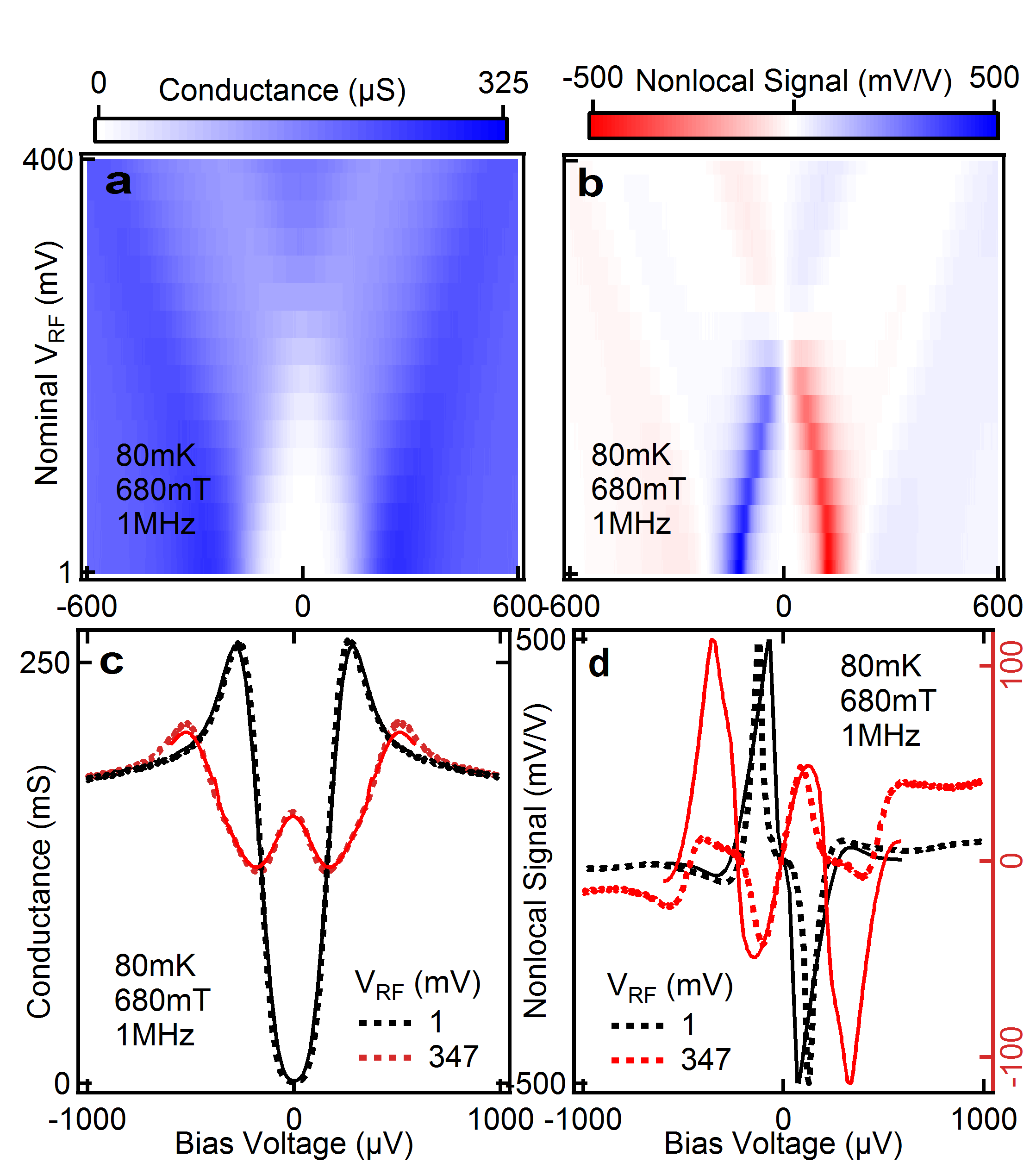}
\caption{\label{fig:mwavepower} (a) Measured local conductance  $dI/dV_{DC}$ across J2 as a function of $V_{RF}$ at $f_{RF}=$ 1MHz and $H=$ 680mT. (b) Measured differential non-local signal $dV_{NL}/dV_{DC}$ at J3 as a function of of $V_{RF}$ at $f_{RF}=$ 1MHz and $H=$ 680mT. Classical rectification is the dominant RF effect. The $V_{RF}$ given here is the value at the output of the generator. As noted in the main text, $V_{RF}$ at the device can be estimated from the classical rectification of features in the $V_{RF}=0$ trace. (c,d) Two slices of (a,b) plotted as dotted lines together with numerical calculations (solid lines) based on the superconducting DOS extracted from the measured local conductance at $V_{RF}=0$. Small `mismatches' in the conductance can lead to large differences in the non-local signal; however, the theory qualitatively agrees with the data.}
\end{figure}

\begin{figure}[!h]
\includegraphics[width=8.6cm]{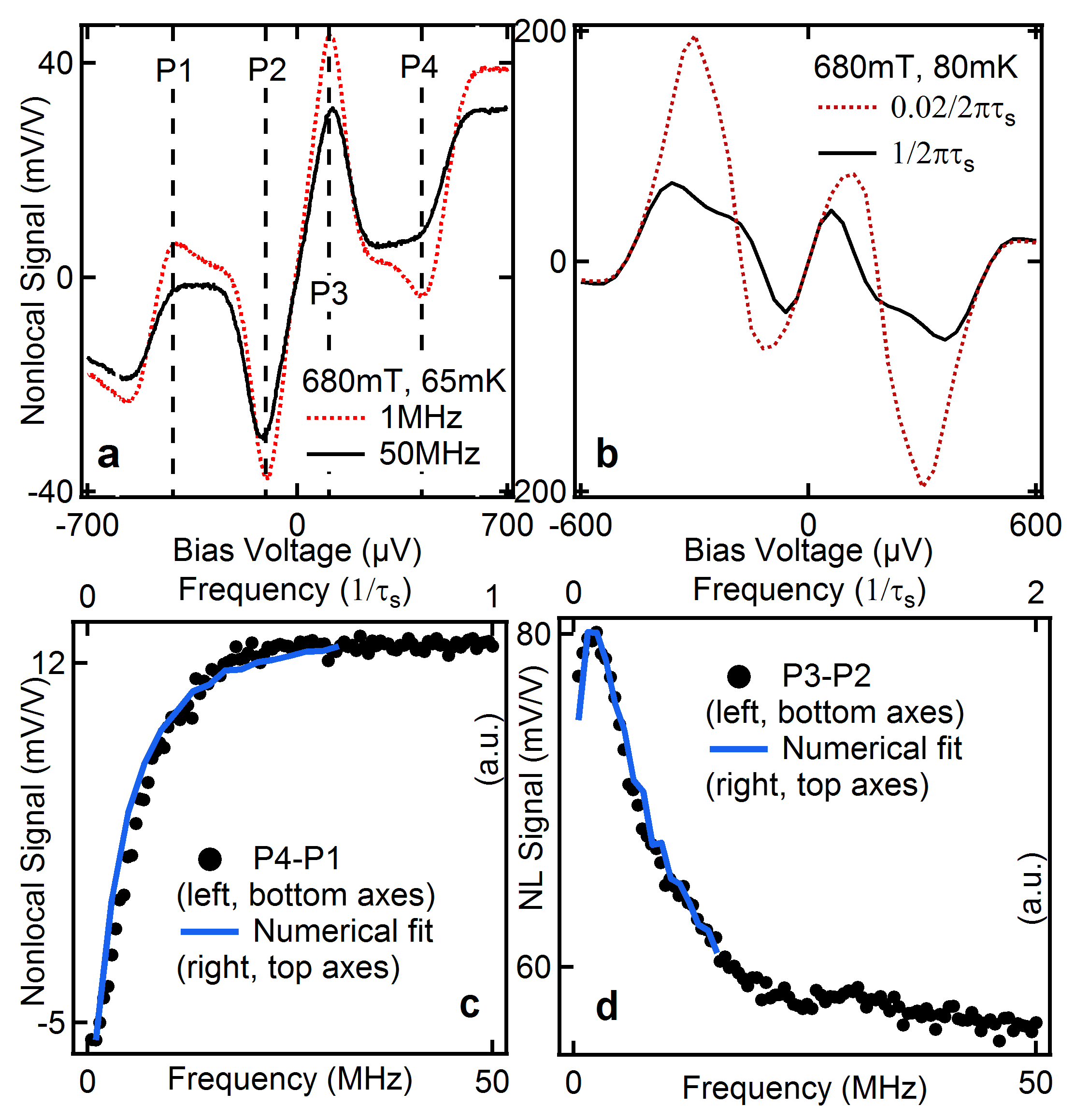}
\caption{\label{fig:cut-off} (a) Differential non-local signal $dV_{NL}/dV_{DC}$ measured at J3 with injection at J2 as a function of $V_{DC}$ with constant-power excitations (of $\sim$250$\mu$V at the device) at $f_{RF}=$ 1MHz, 50MHz. (b) Numerical calculation of $dV_{NL}/dV_{DC}$ as a function of $V_{DC}$ with constant-power excitations at $f_{RF}= 0.02/2\pi\tau_s$, $1/2\pi\tau_s$, based on the superconducting DOS extracted from the measured local conductance at $V_{RF}=0$. (c,d) $dV_{NL}/dV_{DC}$ at the $V_{DC}$ values indicated in (a) as a function of $f_{RF}$. We subtract `opposing' peaks to obtain the anti-symmetric part of the signal, which is due to spin. Fits to numerical calculations yield $\tau_s$ = 3.2ns and 6.4ns with a fitting error of 10-20\%. Note that the cutoff does not occur exactly at $1/(2\pi\tau_s)$.}
\end{figure}

\begin{figure}[!h]
\includegraphics[width=8.6cm]{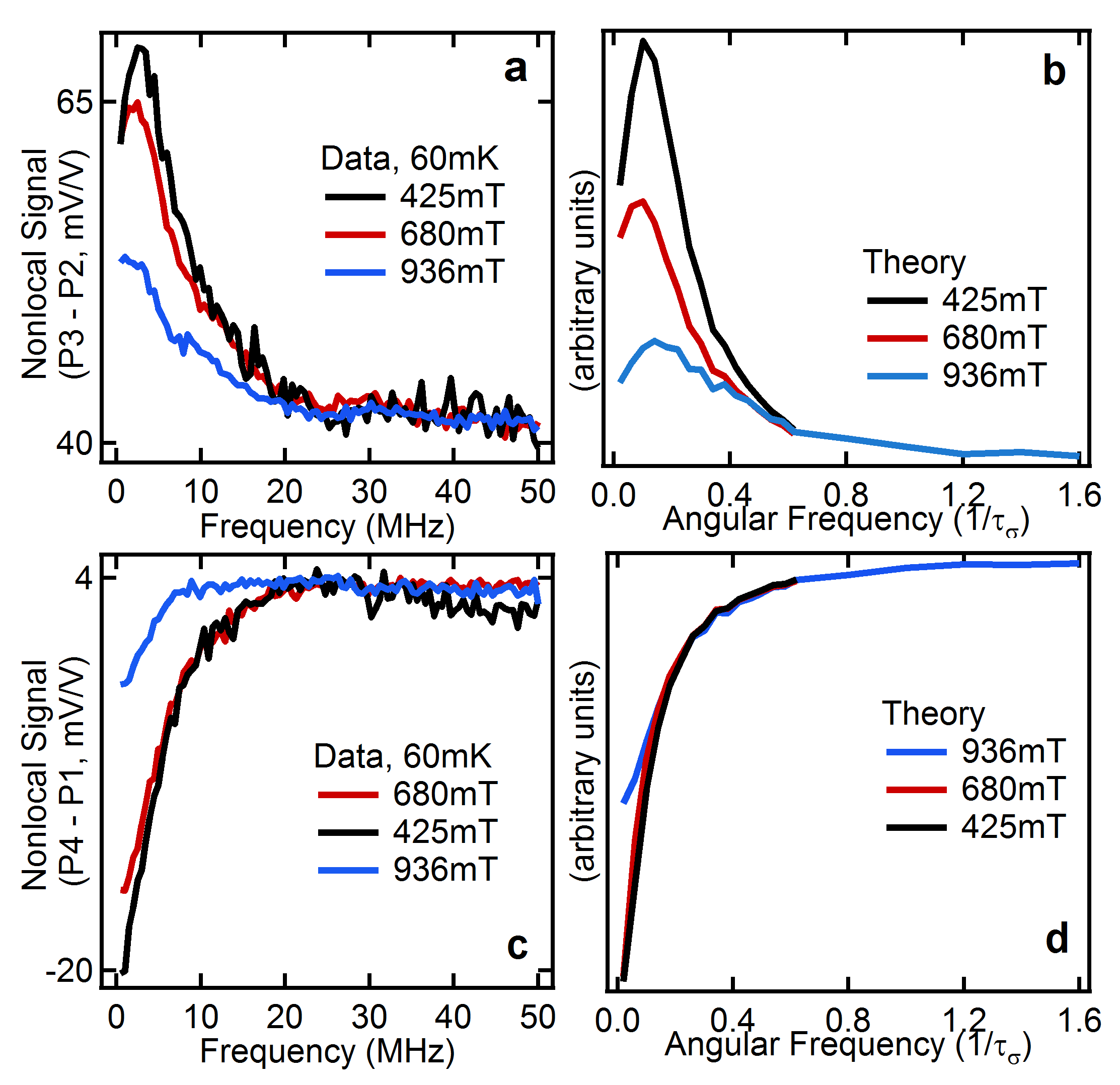}
\caption{\label{fig:fieldtemp} Differential non-local signal $dV_{NL}/dV_{DC}$ as a function of RF frequency at different fields for `inner' and `outer' peaks (cf. Figure 4.), both experimental data (a,c) and theory (b,d). Both show a decrease in amplitude at high fields. Fits of theory to data suggest a slight rise of $\tau_s$ with magnetic field, as well as a difference between inner and outer peaks. (See text.)}
\end{figure}

Turning now to our measurements, we begin by characterising our device in the absence of high-frequency excitation. Figures~\ref{fig:sample}(b) shows the local conductance $dI/dV_{DC}$ as a function of bias voltage and magnetic field. We see that, for this device, the superconducting critical field at J2 is $\sim$1.9T. Figures~\ref{fig:sample}(c) and (d) show the corresponding non-local voltage $V_{NL}$ measured at J1 and J3 respectively. We remind the reader that, as the injection electrode is normal, the (anti-)symmetric part of this signal comes from the spin (charge) imbalance~\cite{quay}. Note that $V_{NL}$ due to spin can be almost half the superconducting gap (Figure~\ref{fig:sample}(d)). As in our previous work, we see a spin signal which first increases with magnetic field then dies out as the magnetic field approaches its critical value  (Figure~\ref{fig:sample}(d)). In contrast, the charge signal diminishes with increasing magnetic field (Figure~\ref{fig:sample}(d) and (e)). Theoretical fits to data at fixed magnetic field such as those shown in Figure~\ref{fig:sample}(e) allow us to estimate the spin lifetime $\tau_S$ at several magnetic fields, yielding 1.2ns, 1.6ns, 2.3ns and 2.6ns for 425mT, 510mT, 680mT and 936mT respectively. Based only on the fits, the error on these figures is about 10-20\%; however, the real value of the error could be larger as it is difficult to theoretically account for orbital depairing effects, due to a small misalignment of the magnetic field with the device plane as well as stray fields from the Co electrode. We emphasise, nevertheless, that our theoretical model is able to reproduce all qualitative features of our data. (Figures~\ref{fig:sample}--\ref{fig:fieldtemp}, Ref.~\cite{chevallier})

At a fixed magnetic field of $H=680mT$ (to obtain a large non-local spin signal), we apply a sinusoidal excitation at 1MHz while sweeping $V_{DC}$ and varying the RF power. (All RF amplitudes given, unless otherwise stated, are those at the output of the generator.) The results are shown in Figure~\ref{fig:mwavepower}. The main effect of the RF excitation on both the local conductance and the non-local signal is the phenomenon known as `classical rectification': As sinusoidal signals spend most time at their extrema, each feature in the original trace is `split' by a distance in bias voltage corresponding to the peak-to-peak amplitude of the RF excitation across the injection junction J2. The splitting of the BCS coherence peaks in the local conductance (Figure~\ref{fig:mwavepower}(a)) as well as that of the spin imbalance peaks in the non-local conductance (Figure~\ref{fig:mwavepower}(b)) are well-reproduced qualitatively by our theory.~(Figure~\ref{fig:mwavepower}(c-d)) (We note that the calculated non-local signal is very sensitive to even small changes in the spin-resolved DOS having almost no effect on the calculated conductance. The local conductance, which we measure, is proportional to the total DOS rather than the spin-resolved DOS.) Figure~\ref{fig:mwavepower} looks the same for all frequencies, modulo an offset in the RF power due to frequency-dependent attenuation in the RF lines. These measurements can thus be used as a calibration of RF power.

Next, we study the frequency-dependent response of our system at constant RF amplitude at the device, using the value of the local conductance at zero bias voltage as a calibration of RF amplitude. (The RF amplitude at the device can be more accurately determined from the location of the `RF-split' peaks and is generally $\sim$250$\mu$V.) Figure~\ref{fig:cut-off}a shows the non-local signal as a function of bias voltage at constant RF amplitude at 1MHz and 50MHz. For both frequencies, `RF-split' peaks appear at the same location, but their amplitudes are different: At frequencies which are high compared to $\sim1/2\pi\tau_S$, the classifically-rectified peaks have smaller amplitudes than they do at low frequencies. (Whether peak amplitudes increase or decrease with frequency depend on the particular parameters of the system~\cite{chevallier}.)

To track the frequency evolution of the peak amplitude, we measure the non-local signal as a function of RF frequency at the bias voltages indicated by the dashed lines (Figures~\ref{fig:cut-off}(c) and (d)). We fit our data to numerical calculations of the peak height to obtain, at 680mT and 60mK, $\tau_s$=6.4ns for the inner peaks and $\tau_s$=3.2ns for the outer peaks. The corresponding figures at 425mT and 936mT are 6.4/3.8ns and 8/8.6ns for inner/outer peaks, with a fitting error of 10-20\%. As also observed in~\cite{wolf}, our data show no changes in $\tau_s$ with increasing temperature up to 600mK as the quasiparticle population is strongly out-of-equilibrium. (See Supp. Info.)

Note that both the inner and outer RF-split peaks originate from the same spin-imbalance peak; however, in the case of the inner (outer) the original peak is `excited' together with quasiparticles of lower (higher) energy. Our results thus suggest that $\tau_s$ may depend on the quasiparticle distribution, but this conclusion can only be tentative at this juncture due to the sensitivity of the calculated non-local signal to the spin-resolved DOS. Thus, while our experimental techniques open up the possibility of studying the distribution dependence of $\tau_s$ (which should insights into the role of quasiparticle-quasiparticle interactions on spin relaxation), further progress on both theoretical and experimental fronts are needed: On the theoretical end, more accurate calculations of the spin-resolved DOS could be attempted, while on the experimental end, the (stray) out-of-plane field could be minimised.

The results of measurements similar to those shown in Figure~\ref{fig:cut-off}, performed at different fields and at the base temperature of the dilution refrigerator ($\sim$60mK) are shown in Figure~\ref{fig:fieldtemp} together with numerical calculations. No significant change in the cut-off frequency (within measurement error) was observed in the range of magnetic fields; however, as mentioned above, the numerical fits suggest a slight rise in $\tau_s$ with increasing magnetic field. This rise is smaller than that measured in our previous work on thicker Al samples (20nm vs 8.5nm here); this is consistent with results by the Karlsruhe group~\cite{wolf}, which also indicates a flatter field dependence for thinner samples.

Finally, the $\tau_s$ we obtain both from the fits to the DC data and from the frequency cut-offs are lower than those obtained in our previous work~\cite{quay}. As also observed by the Karlsruhe group~\cite{wolf} $\tau_s$ decreases for thinner films.There could be several physical reasons for this, including increased scattering (lower mean free path) and therefore increased spin-flip scattering~\cite{beuneu,meservey-review}. Although the increased importance of spin-orbit effects at the surface~\cite{meservey-review} and finite size effects~\cite{long} give the right qualitative thickness dependence for $\tau_s$, the spin-orbit scattering measured in Al thin films of the same thickness is two order of magnitude smaller ($ \approx 50ps$) ~\cite{quay-qsr}. Both our results and those in Ref. ~\cite{wolf} are consistent with $\tau_s \approx \tau_{ee}$ where $\tau_{ee}$ is the electron-electron scattering time. Using the expression for enhanced electron-electron scattering time in thin films predicted by Abrahams-Anderson-Lee-Ramakrishnan~\cite{abrahams} we obtain  $\tau_{ee} \approx$ 5ns for our samples ($R_{sq}=14 \Omega$ and $T=60mK$) and $\tau_{ee} \approx 15 ns $ for thicker Al films in Ref.~\cite{wolf}, consistent with previous works ~\cite{santhanam}, ~\cite{vanson}. To verify this estimate, we have measured $\tau_{ee}$ in the frequency domain following the method presented in Ref. ~\cite{vanson}, which is based on the enhancement of the critical pair-breaking current by microwave radiation --- we obtain $\tau_{ee} \approx 3 ns $ (See Supp. Info.).

Note that the measured cut-offs are independent of the value of the coupling capacitance of the RF line and of the detector's differential resistance at $V_{DC}=0$ (due to different levels of depairing due to stray fields), thus ruling out detector bandwidth effects. We also checked that the injection of electrons at several times the superconducting gap energy did not significantly affect the shape of the coherence peaks, and the superconducting gap changes by $\approx 2\%$ at most. (See Supp. Info.)

In conclusion, we have measured the lifetime of spin imbalances in the quasiparticle population of a superconductor in the frequency domain. This is the most direct measurement to date of this quantity and our technique enables the study of the role of quasiparticle-quasiparticle interactions in spin relaxation. The charge lifetime could in principle be measured in a similar way, at much higher excitation frequencies. Pushing these experiments one step further, one could look at variations in the spin accumulation either in real-time or at the excitation frequency. All of these techniques could in principle be used to measure spin lifetimes in other superconducting materials in the Pauli limit.

\begin{acknowledgments}

We thank J. Gabelli for helpful discussions on spin dynamics in superconductors, and J. S. Meyer, M. Houzet and T. Krishtop for the same on effective-temperature-induced spin imbalances. This work was funded by European Research Council Starting Independent Researcher (NANO-GRAPHENE 256965) and Synergy Grants; an ANR Blanc grant (MASH) from the French Agence Nationale de Recherche; and the Netherlands Organization for Scientific Research (NWO/OCW).
\end{acknowledgments}

\section{Supplementary Information}

\subsection{Measurement Circuit}

\begin{figure}[H]
\centering
\includegraphics[width=11cm]{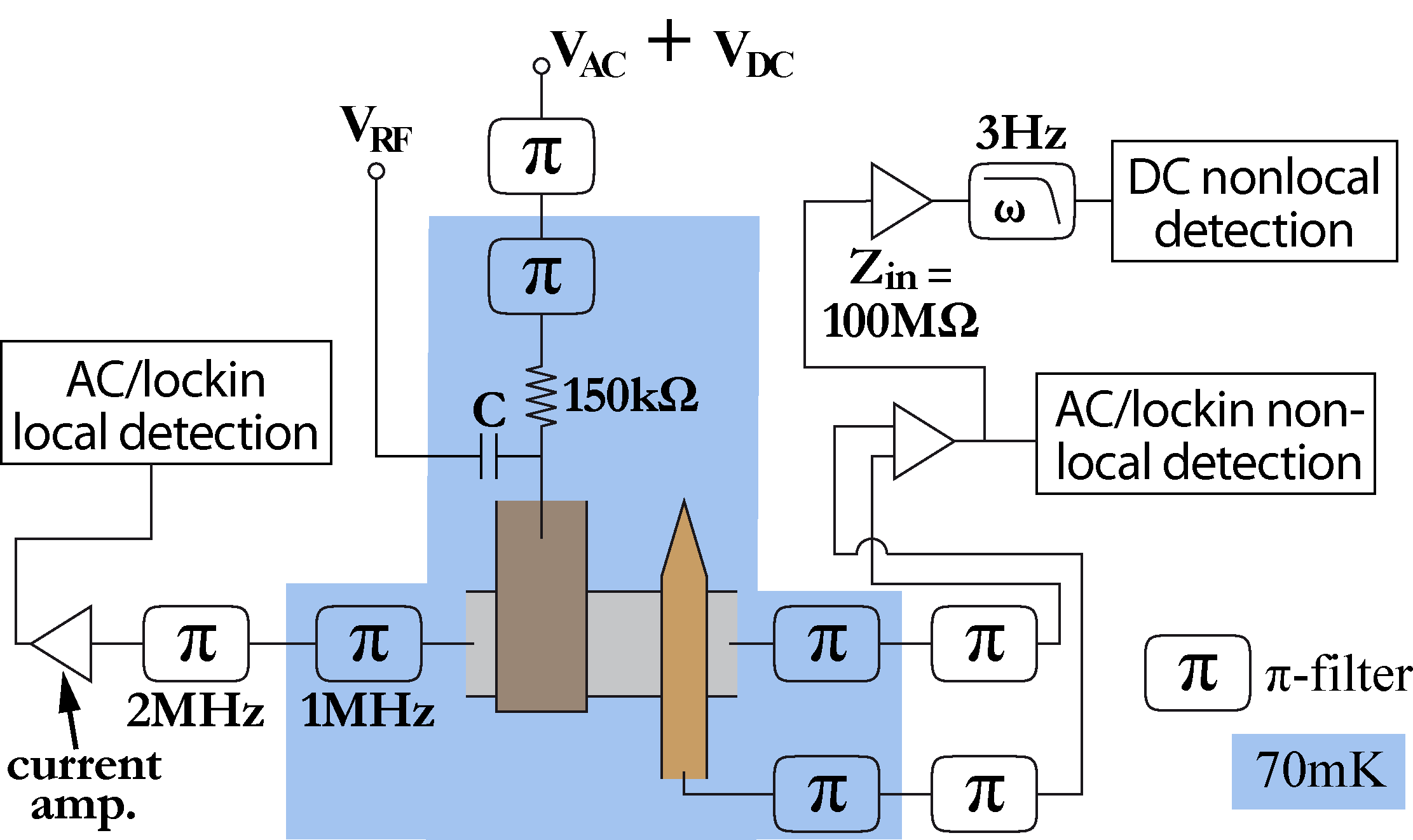}
\caption{\label{fig:circuit} Detailed diagram of the measurement circuit used in the experiment.}
\end{figure}

Figure~\ref{fig:circuit} shows our measurement circuit in greater detail than was presented in the main text. All voltage amplifiers have input impedances of 100M$\Omega$. All $\pi$-filters at low temperature have cutoff frequencies of 1MHz while those at room temperature have cutoff frequencies of 2MHz.

The two lockin measurements are synchronized at 7Hz or 9Hz, the AC excitation frequency. The amplitude of the AC excitation $V_{AC}$ is 10$\mu$V for Figures 2 and 3 of the main text. For Figures 4 and 5, $V_{AC}$ was increased to 20$\mu$V to improve the signal-to-noise ratio of the frequency cut-off traces. This only minimally affected the shape of the traces as they no longer contained sharp features.

The value of the capacitor denoted $C$ was 1.5nF for the measurements shown in the main text and 15nF for measurements done at other points. In very early measurements, there was a bias-tee in the circuit instead of the resistor and capacitor shown. The observation of the frequency cut-off does not depend on the specifics of the circuit.

\subsection{Injecting from J1}

\begin{figure}[H]
\centering
\includegraphics[width=11cm]{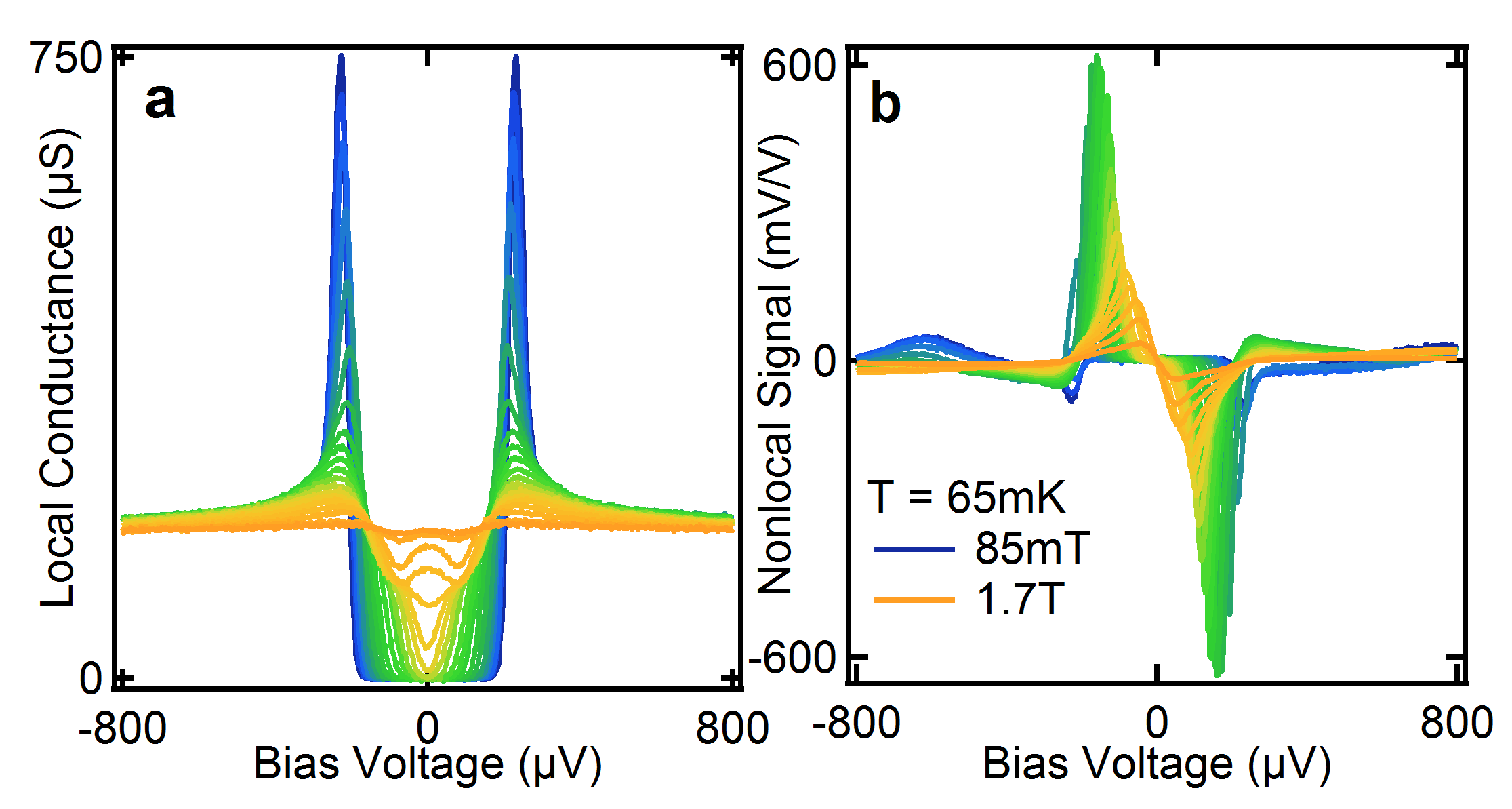}
\caption{\label{fig:j1injection} (a) Local conductance $dI/dV_{DC}$ measured at J1, over a range of magnetic fields. (b) Corresponding differential non-local signal $dV_{NL}/dV_{DC}$ measured at J3.}
\end{figure}

As noted in the main text, all the data shown were with quasiparticles injection into the superconductor at the junction J2 and detection at J3.

In Figure~\ref{fig:j1injection}, we show data from the same device, with injection across J1 instead and detection still at J3. The measured local conductance, proportional to the density of states (DOS) in the superconductor (S), shows that the latter is less depaired at J1 than at J2 (compare Figure~\ref{fig:j1injection}a to Figure 2a of the main text). This is because the main cause of depairing is stray fields from the cobalt electrode at J3, which can have a component perpendicular to the plane of the device. As the distance J1--J3 is larger than J2--J3, the S DOS is less depaired at J1 compared to J2. Thus, the Zeeman splitting of the DOS at high magnetic fields is also more apparent in these data. The non-local differential signal can also be seen to be larger (Figure~\ref{fig:j1injection}b) (To a very rough first approximation, it is proportional to the derivative of the injection DOS~\cite{quay}. While this is no longer true in this case, generally the `sharper' the injection and detection DOS, the larger the non-local signal.)

N.B. The data shown in Figure~\ref{fig:j1injection}a are the in-phase component of the differential non-local signal. As J3 is very resistive ($e\mu_s\sim\Delta/2$ at most and the density of states at J3 highly non-linear, see Figure~\ref{fig:cobaltdos}), the out-of-phase component can be of comparable amplitude; however, we checked that the in-phase signal is the same in all its essential details to the numerical derivative (with respect to the bias voltage) of the DC non-local signal to an overall factor of $\lesssim2$.

\section{Out-of-Equilibrium State of the System}

\begin{figure}[H]
\centering
\includegraphics[width=16.5cm]{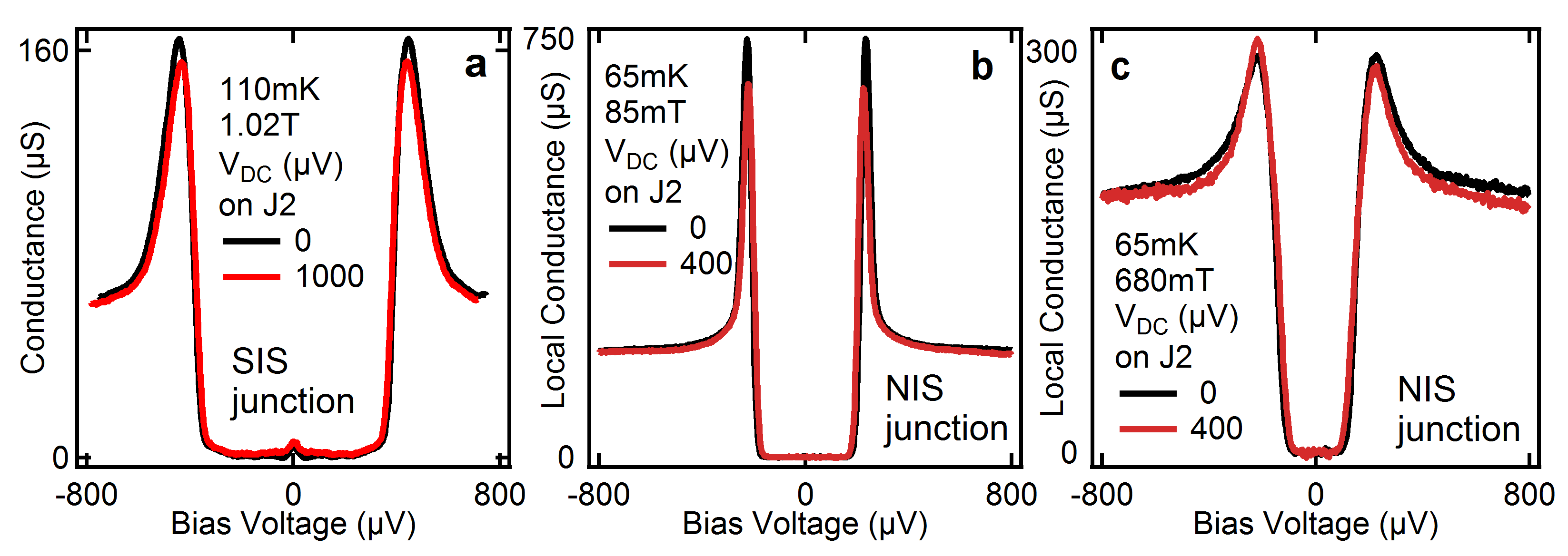}
\caption{\label{fig:thermoelectric} Local conductance $dI/dV_{DC}$ at J1 (`detector') with and without a large voltage bias on J2 (`injector') for two kinds of detector junctions: (a) superconductor-insulator-superconductor and (b,c) normal-insulator-superconductor as in the main text.}
\end{figure}

Figure~\ref{fig:thermoelectric}a shows data from a similar to that shown in the main text, but where the electrode at J1 instead of being normal is superconducting. (The J1-J2 distance here is 1.8$\mu$m) We measured the conductance at J1 while biasing J2 very much above the superconducting gap, thus creating large spin and charge imbalances at J1. We compare this to the same measurement when J2 is not biased. The conductance across an SIS junction can be broadened by an increase in the quasiparticle temperature in either superconductor. It can be seen that biasing J2 does not do much to broaden the conductance at J1, while the gap is decreased by $\approx 2\%$.

Biasing J2 in the devices shown in the main text also induces a slight narrowing of the gap at J1, which seems similarly not to be due to an effective temperature increase: We also performed similar measurements on the devices presented in the main text. (Figures ~\ref{fig:thermoelectric}b and c.) The same narrowing of the gap at J1 when J2 is biased far above the gap is $\approx4\%$ at $H=85mT$ as shown in Figure~\ref{fig:thermoelectric}b. (The conductance across a NIS junction is not sensitive to the temperature of the quasiparticles in the superconductor. The asymmetry in the coherence peaks we have determined to be an artefact.)

Thus we see that, in all of the measurements described in the main text, the shape of the BCS coherence peaks in the superconductor is unaffected by the injection, and that the gap decreases by $2\%$ at most. As the current-voltage characteristics of the junctions are highly non-linear, this is an over-estimate.

\subsection{RF Amplitude Calibration}

\begin{figure}[H]
\centering
\includegraphics[width=11cm]{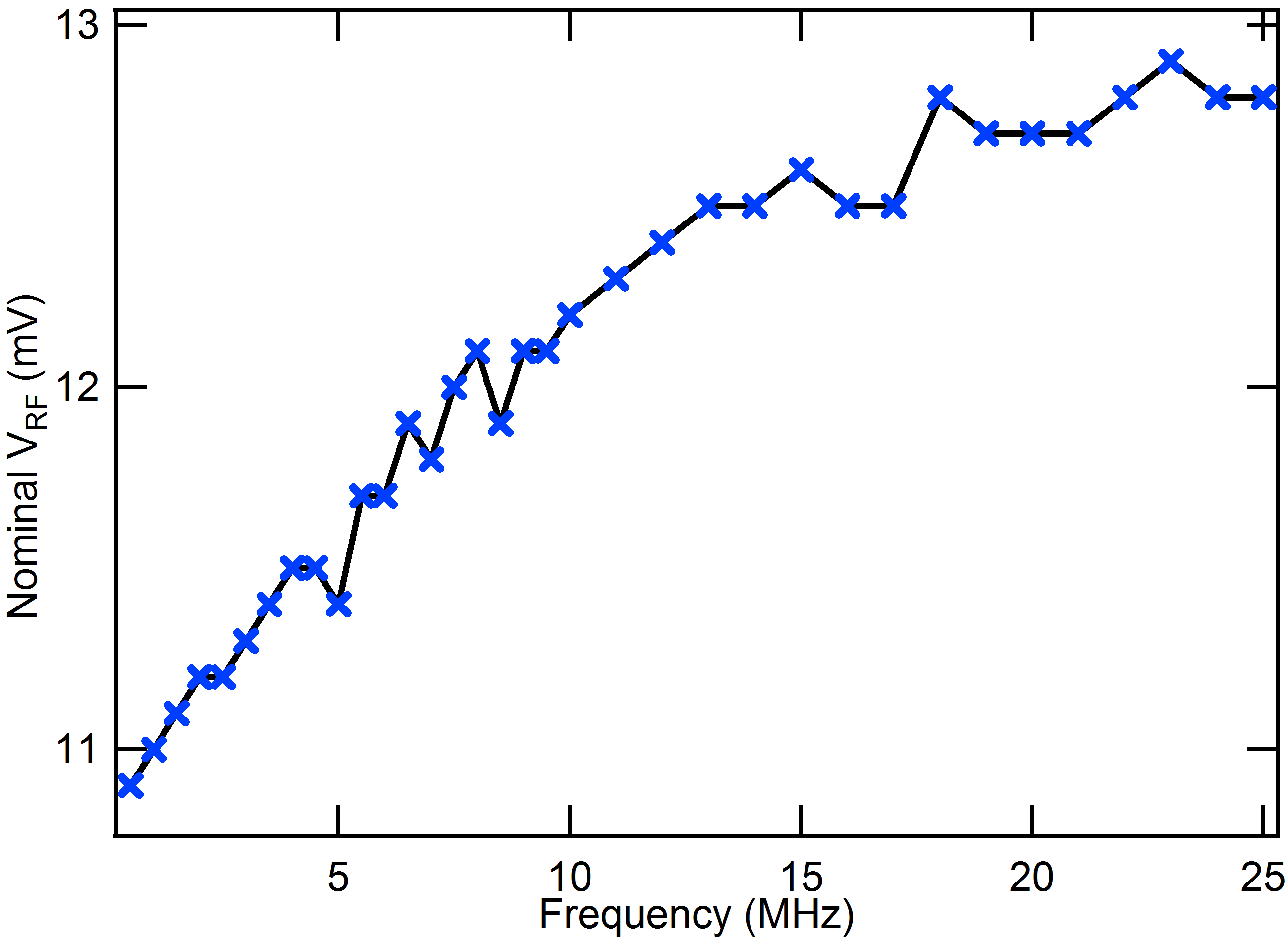}
\caption{\label{fig:freqcalib} A typical RF amplitude ($V_{RF}$) calibration trace. As a function of frequency, the nominal $V_{RF}$ required from the RF generator in order to obtain a constant $V_{RF}$ measured at the device of about 256$\mu$V.}
\end{figure}

As mentioned in the main text, we use local conductance measurements at the injection junction to calibrate the amplitude of the RF signal $V_{RF}$ arriving at the sample. The locations of peaks in the differential non-local signal at the detector junction provide additional confirmation that $V_{RF}$ is constant as we change the frequency.

Figure~\ref{fig:freqcalib} shows the nominal $V_{RF}$ value: the RF amplitude at the output of the generator. Room temperature measurements show that our RF line attenuates more at higher frequencies; thus the generator amplitude has to increase to compensate. The nominal $V_{RF}$ shown here result in a $V_{RF}$ at the device of 256$\mu$V as measured by the distance between `RF-split' peaks in the differential non-local signal.

The data shown were taken with an attenuator of -40dB (specified for 50$\Omega$) between the generator and our RF line. Note that, as the impedance of our device is high compared to 50$\Omega$, the attenuator is less effective in our measurement than its nominal value (it attenuates by less than -40dB). In addition, the high impedance boundary condition at the end of the RF line can also `increase' $V_{RF}$ at the device.

\subsection{Very High-Frequency Injection}

\begin{figure}[H]
\centering
\includegraphics[width=6cm]{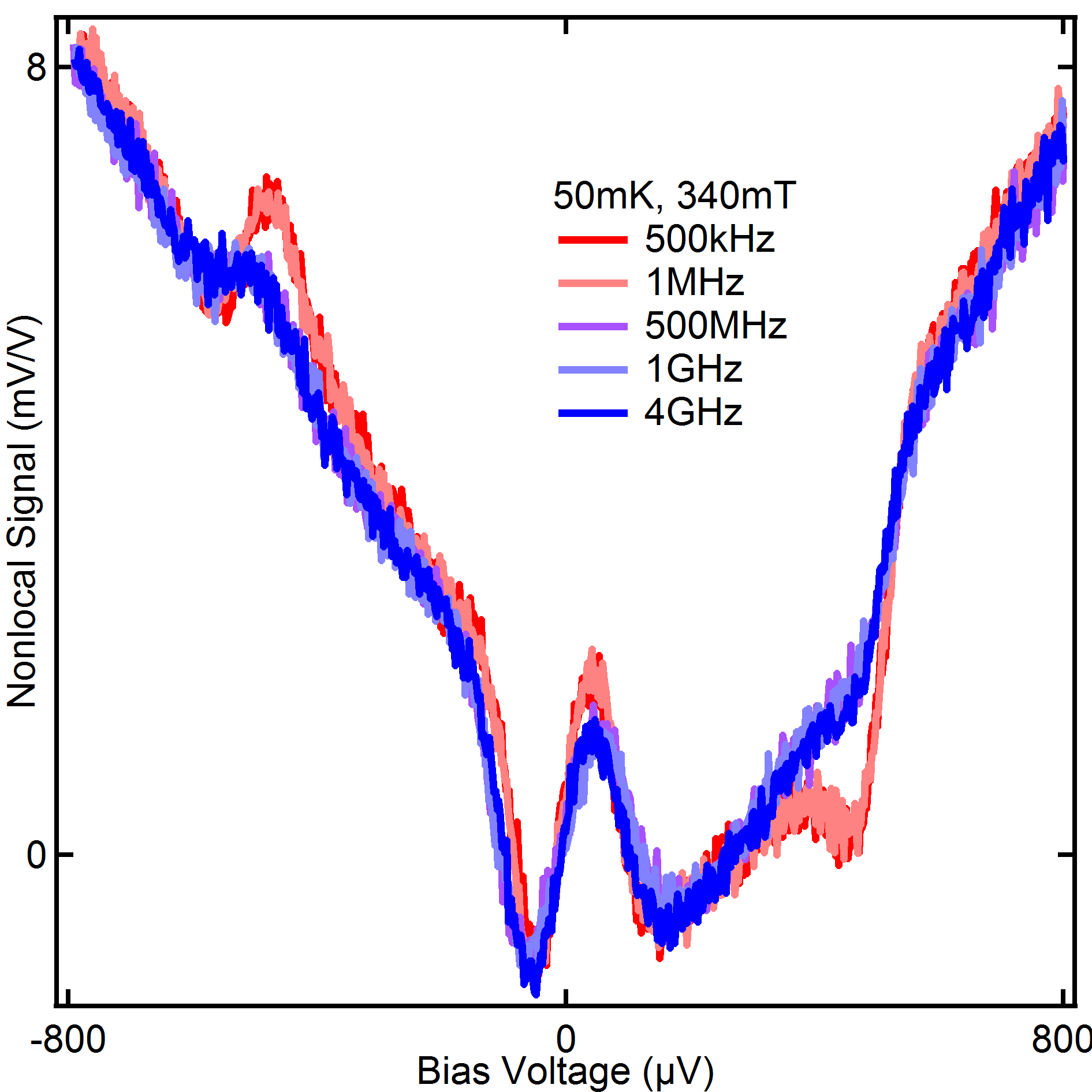}
\caption{\label{fig:hifreq} Differential non-local signal $dV_{NL}/dV_{DC}$ as a function of $V_{DC}$ at different $f_{RF}$. These data are from a device which is different from, but nominally identical to, the one from which data are shown in the main text.}
\end{figure}

To verify that the behaviour we observe in the frequency domain is indeed a cut-off rather than e.g. an oscillatory phenomenon, we also performed measurements at RF injection frequencies very much above the observed cut-off. Figure~\ref{fig:hifreq} shows the differential non-local signal $dV_{NL}/dV_{DC}$ as a function of DC bias voltage $V_{DC}$ at several RF injection frequencies (and at constant amplitude $V_{RF}$). Between 500kHz and 500MHz, similar to data shown in the main text, the amplitudes of the classically-rectified spin imbalance peaks diminishes. However, there is no further change in the signal at higher frequencies, in agreement with our theoretical expectation of a cut-off.

\subsection{Cobalt Polarisation}

\begin{figure}[H]
\centering
\includegraphics[width=9cm]{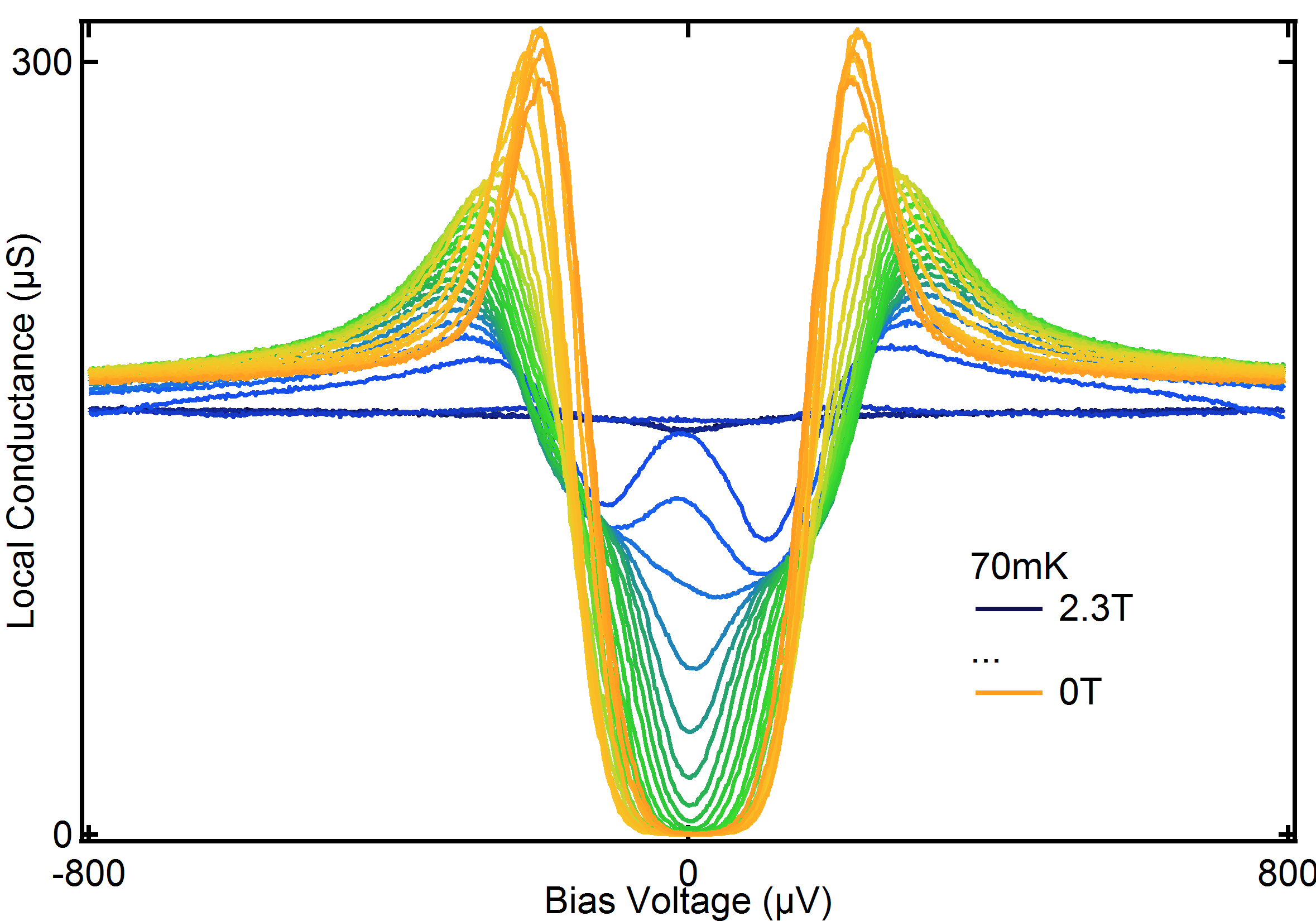}
\caption{\label{fig:cobaltdos} Local conductance measured at J3, over a range of magnetic fields.}
\end{figure}

In order to estimate the polarisation, $P$ of conduction electrons at the Fermi level of the cobalt electrode and to determine its sign, we measure local conductance spectra at J3 as a function of magnetic field. (Figure~\ref{fig:cobaltdos}) The bias voltage here is applied between the Co (positive voltage) and Al (ground) electrodes. From the asymmetry of the inner Zeeman-split peaks, it is already possible to see that magnetic moments in the Co are polarised in the direction of the applied field (cf. Ref.~\cite{tedrow,fulde}). (Note that electron spins are anti-aligned to their magnetic moments as in Ref.~\cite{tedrow}.) From the heights of the inner `shoulders' we estimate $P$ to be 7--10\%~\cite{tedrow,paraskevopoulos}, consistent with results from previous work on similar samples~\cite{quay}.

\subsection{Temperature Dependence}

\begin{figure}[H]
\centering
\includegraphics[width=6cm]{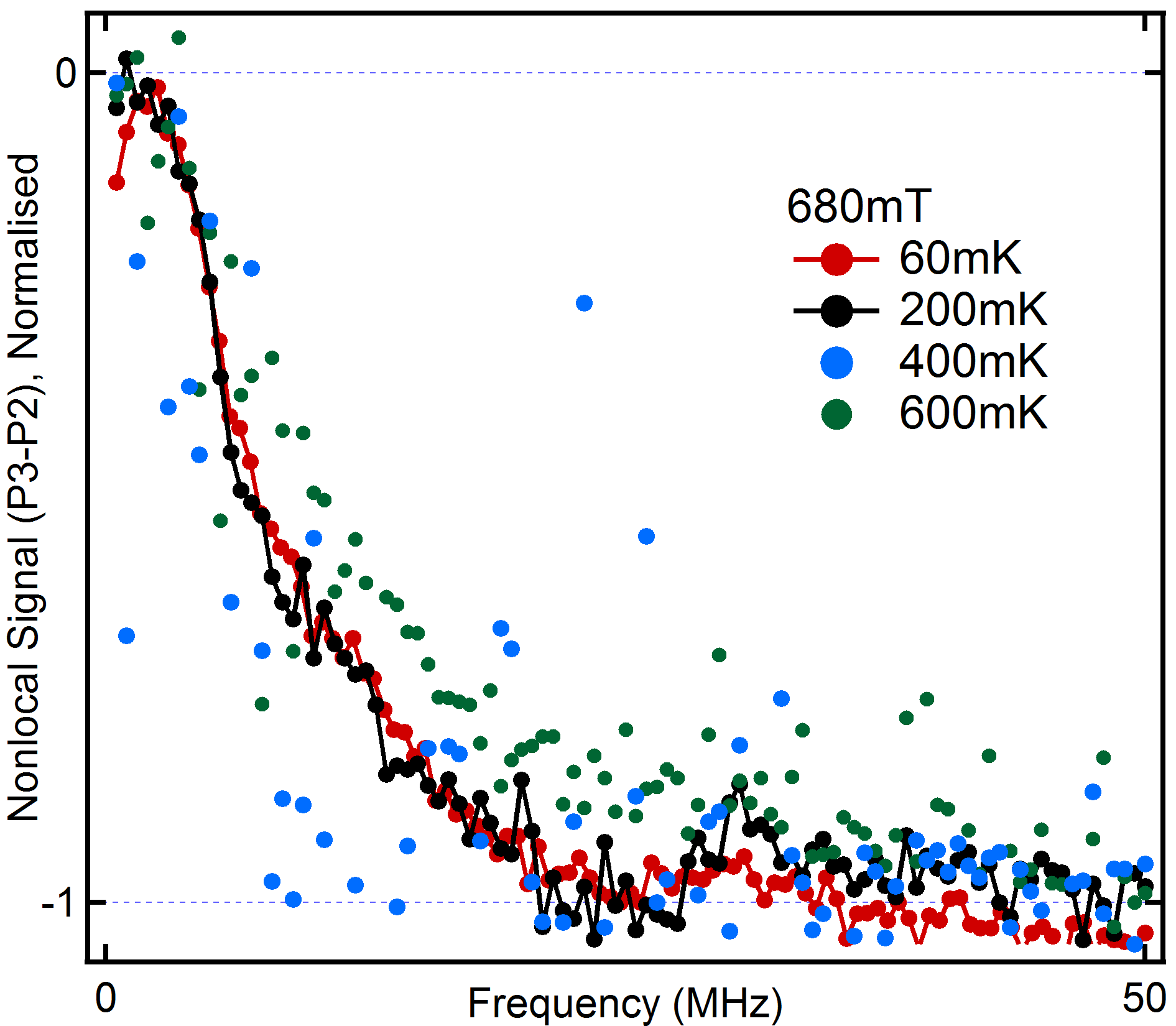}
\caption{\label{fig:temperature} Normalised differential non-local signal $dV_{NL}/dV_{DC}$ measured at J3 with injection at J2 as a function of $f_{RF}=$ and temperature at a fixed magnetic field of 680mT. As in Figures 3 and 4 of the main text, values at `opposing' peaks are subtracted.}
\end{figure}

Figure~\ref{fig:temperature} shows data similar to that in Figure 3d of the main text, at different temperatures. The traces have been normalised for clarity. Within the limits of the measurement, there appears to be no significant dependence of the visible cutoff frequency on temperature.

\subsection{Frequency Domain Measurement of $\tau{ee}$}

\begin{figure}[H]
\centering
\includegraphics[width=17cm]{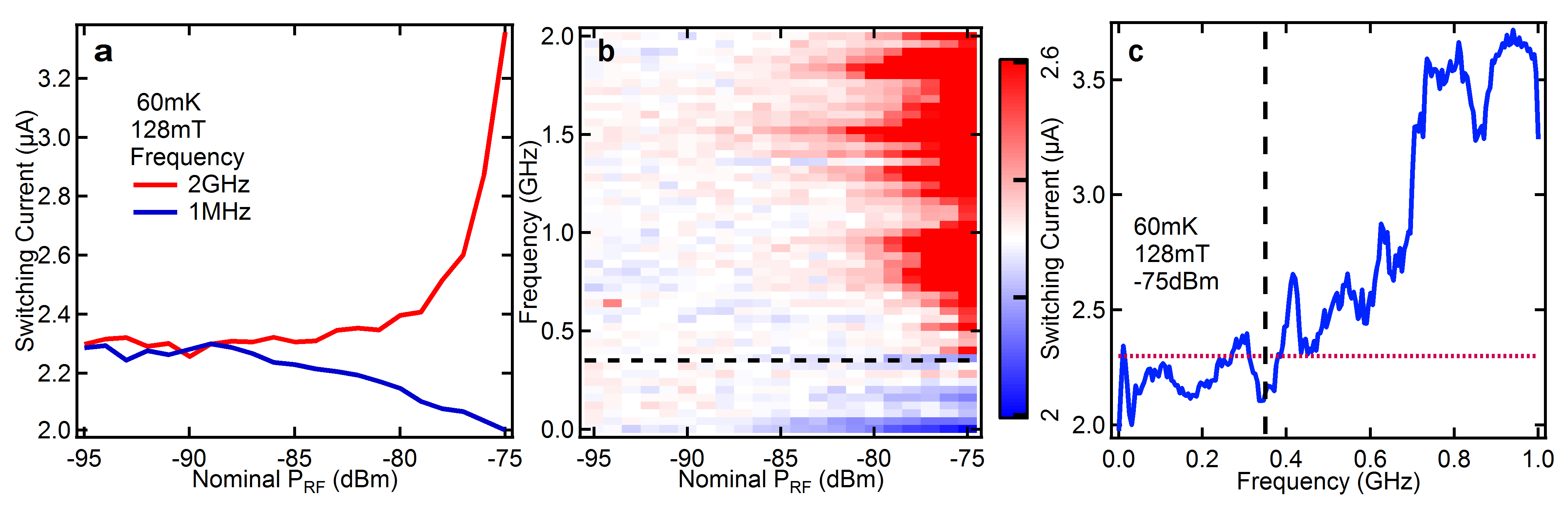}
\caption{\label{fig:tau_ee} (a,b) Switching current of the superconducting Al bar, in a device similar to that shown in the main text, as a function the frequency and power of incident microwave radiation. The dashed black line in (b) shows the frequency corresponding to $\tau_{ee}$, the electron-electron interaction time. (c) Switching current of the same device as a function of frequency at a constant nominal microwave power of -75dBm. The dotted red line shows the switching current in the absence of microwaves. The oscillations in the signal are due to resonances in the coaxial cable used to transmit the microwaves to the device.}
\end{figure}

We measure $\tau_{ee}$ in the frequency domain following the method presented in Ref. ~\cite{vanson}. We measure the switching current, $I_S$ of a 6nm think Al bar, in a device similar to that used for the measurements in the main text. At the same time, a sinusoidal (microwave) signal of frequency $f_{RF}$ is applied across the bar via a lossy coaxial cable and a bias tee at low temperature. Figure~\ref{fig:tau_ee} show $I_S$ as a function of $f_{RF}$ and microwave power. (The power reported is that at the output of the high-frequency generator.)

At low frequencies, the microwaves cause pair-breaking (as does a DC current) and reduces $I_S$. At high frequencies, enhancement of the superconducting gap occurs due to quasiparticle population being driven out-of-equilibrium by the microwaves. In Ref. ~\cite{vanson}, we see that the crossover occurs at $f_{RF}^c \sim \tau_{ee}$. We measure $f_{RF}^c \sim $ 350MHz, which corresponds to a $\tau_{ee}$ of $\sim$ 3ns.

\end{document}